# INCLUSIVE HADRONIC PRODUCTION OF THE $B_c$ MESON VIA HEAVY QUARK FRAGMENTATION [*]

KINGMAN CHEUNG[†]

*Center for Particle Physics, University of Texas at Austin, Austin TX 78712, U.S.A.*
*E-mail: cheung@utpapa.ph.utexas.edu*

## ABSTRACT

We summarize the studies on the hadronic production of S- and P-wave $(\bar{b}c)$ mesons via direct fragmentation of the bottom antiquark as well as the Altarelli-Parisi induced gluon fragmentation.

The direct production of heavy mesons like $J/\psi$, $\Upsilon$, and $(\bar{b}c)$ mesons can provide very interesting tests for perturbative QCD. According to the potential model calculation [1], for $(\bar{b}c)$ mesons the first two sets ($n=1$ and $n=2$) of S-wave states, the first ($n=1$) and probably the entire second set ($n=2$) of P-wave states, and the first set ($n=1$) of D-wave states lie below the $BD$ flavor threshold. Since the annihilation channel of excited $(\bar{b}c)$ mesons is suppressed relative to the electromagnetic and hadronic transitions, the excited states below the $BD$ threshold will cascade down into the ground state $B_c$ via emission of photons and/or pions. Inclusive production of the $B_c$ meson therefore includes the production of the $n=1$ and $n=2$ S-wave and P-wave states, and the $n=1$ D-wave states. Here we do not include the D-wave contributions since they are expected to be very small.

Intuitively, the dominant production of $(\bar{b}c)$ mesons at the large transverse momentum region should come from the direct fragmentation of the heavy $\bar{b}$ antiquark [2,3]. We calculate the hadronic production of S- and P-wave $(\bar{b}c)$ mesons using the fragmentation approach [4,5,6]. The fragmentation approach essentially involves the factorization of the whole production process into the production of a high energy parton (a $\bar{b}$ antiquark or a gluon) and the fragmentation of this parton into various $(\bar{b}c)$ states. The novel feature in our approach [2,3] is that the relevant fragmentation functions at the heavy quark mass scale can be calculated in perturbative QCD. Let $H$ denotes any $(\bar{b}c)$ meson states. The differential cross section $d\sigma/dp_T$ versus the transverse momentum $p_T$ of $H$ is given by

$$\frac{d\sigma}{dp_T}(p\bar{p}\to H(p_T)X) = \sum_{ij}\int dx_1 dx_2 dz\, f_{i/p}(x_1,\mu) f_{j/\bar{p}}(x_2,\mu) \left[\frac{d\hat\sigma}{dp_T}(ij\to \bar{b}(p_T/z)X,\mu)\right.$$
$$\left.\times D_{\bar{b}\to H}(z,\mu) + \frac{d\hat\sigma}{dp_T}(ij\to g(p_T/z)X,\mu)\, D_{g\to H}(z,\mu)\right]\,. \quad (1)$$

For the production of $\bar{b}$ we include the subprocesses $gg\to b\bar{b}$, $g\bar{b}\to g\bar{b}$, and $q\bar{q}\to b\bar{b}$; while for the gluon $g$ we include the subprocesses $gg\to gg$, $q\bar{q}\to gg$, and $gq(\bar{q})\to$ $gq(\bar{q})$. In Eq. (1), a common scale $\mu$ is chosen for parton distribution functions, parton-parton scattering, and fragmentation functions. We estimate the dependence on $\mu$ by varying the scale $\mu = (0.5 - 2)\mu_R$, where $\mu_R = \sqrt{p_T^2(\text{parton}) + m_b^2}$. This choice of scale avoids the large logarithms in the short-distance part $\hat{\sigma}$'s. However, logarithms of order $\mu_R/m_b$ have to be summed in the fragmentation functions, which is implemented by evolving the Altarelli-Parisi (AP) equations for the fragmentation functions [4,5]. The initial conditions for the AP equations are the fragmentation functions that we can calculate by perturbative QCD at the initial scale $\mu_0$, which is of the order of the $b$-quark mass. At present, all the S-wave [2] and P-wave [3] fragmentation functions for $\bar{b} \to (\bar{b}c)$ have been calculated to leading order in $\alpha_s$. To obtain the fragmentation functions at an arbitrary scale greater than $\mu_0$, we numerically integrate the AP evolution equations.

---
[*]Talk presented at Beyond the Standard Model IV, Lake Tahoe, California (Dec 1994)
[†]Representing also Tzu Chiang Yuan, UC-Davis

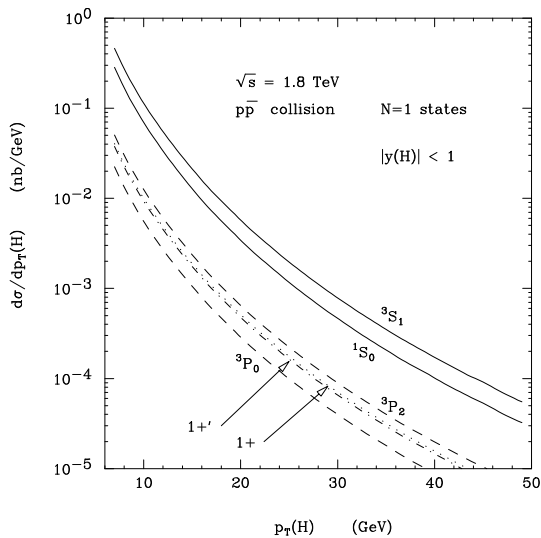

Figure 1: The differential cross section $d\sigma/p_T$ versus $p_T$ of the $(\bar{b}c)$ meson $(H)$ in various spin-orbital states with $n = 1$ at the Tevatron. The acceptance cuts are $p_T(H) > 6$ GeV and $|y(H)| < 1$.

Other details in inputs can be found in Ref. 6. We impose $p_T(H) > 6$ GeV and $|y(H)| < 1$ cuts on the $(\bar{b}c)$ state $H$. The $p_T$ spectra for the $(\bar{b}c)$ state $H$ with various spin-orbital quantum numbers are shown in Fig. 1 and Fig. 2 for $n = 1$ and $n = 2$, respectively. Thus, we can also obtain the inclusive production rate of $B_c$ as a function of $p_T^{\min}(B_c)$ by integrating the $p_T$ spectra. Table 1 gives the inclusive cross sections for the $B_c$ meson at the Tevatron as a function of $p_T^{\min}(B_c)$, including $n = 1$ and $n = 2$ S- and P-wave state contributions. The variations versus the scale $\mu$ between $\mu_R/2$ and $2\mu_R$ are always within a factor of two, and are rather insensitive to changes in scale when $p_T^{\min}(B_c) \gtrsim 10$ GeV.

At the end of Run Ib at the Tevatron, the total accumulated luminosities can be up to 100–150 pb$^{-1}$ or more. With $p_T > 6$ GeV, there are about $5 \times 10^5$ $B_c^+$ mesons.

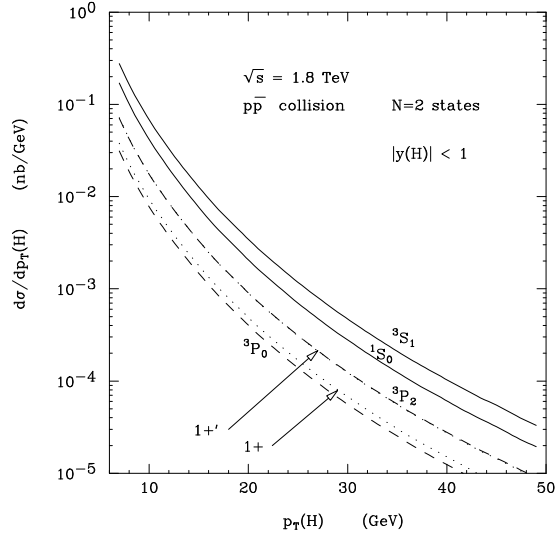

Figure 2: Same as Fig.1 for $n = 2$.

In the future, when the Main Injector is installed in the Run II, which can accumulate 1–2 fb$^{-1}$ luminosity, there will be of order $10^7$ $B_c$ mesons. At the LHC there will be about $3 \times 10^9$ $B_c$ mesons with $p_T > 10$ GeV at the assumed 100 fb$^{-1}$ luminosity.

In conclusion, there should be enough signature events to confirm the existence of $B_c$ at the Tevatron, and the LHC will be a copious source of $B_c$. This work was supported by US DOE-FG03-93ER40757.

1. E. Eichten and C. Quigg, Phys. Rev. **D49**, 5845 (1994).
2. E. Braaten, K. Cheung, and T. C. Yuan, Phys. Rev. **D48**, R5049 (1993).
3. T. C. Yuan, Phys. Rev. **D50**, 5664 (1994).
4. K. Cheung, Phys. Rev. Lett. **71**, 3413 (1993).
5. K. Cheung and T. C. Yuan, Phys. Lett. **B325**, 481 (1994).
6. K. Cheung and T.C. Yuan, preprint UCD-95-4 and CPP-94-37 (Feb 1995).

Table 1: Inclusive production cross sections for the $B_c$ meson at the Tevatron including the contributions from all the S-wave and P-wave states below the $BD$ threshold as a function of $p_T^{\min}(B_c)$. The acceptance cuts are $p_T(B_c) > 6$ GeV and $|y(B_c)| < 1$.

| $p_T^{\min}$ (GeV) | $\sigma$ (nb) | | |
|---|---|---|---|
| | $\mu = \tfrac{1}{2}\mu_R$ | $\mu = \mu_R$ | $\mu = 2\mu_R$ |
| 6 | 2.81 | 5.43 | 6.93 |
| 10 | 0.87 | 1.16 | 1.22 |
| 15 | 0.26 | 0.29 | 0.26 |
| 20 | 0.098 | 0.097 | 0.083 |
| 30 | 0.021 | 0.018 | 0.014 |